# Thermodynamic analysis of water sorption isotherms on magnetically treated maize (*Zea mays* L.) seeds.

**Iván D. Buitrago-Torres, Javier I. Torres-Osorio**

## Abstract

Magnetic seed treatment (MST) is an effective technique for improving conditions in processes associated with germination, stability, and, ultimately, crop production. In this work, the technical management parameters related to the water adsorption process on magnetically treated Zea mays L. seeds will be determined. Models were compared to describe the experimentally obtained adsorption isotherms using nonlinear regressions. The modeling parameters, along with the Clausyus-Clapeyron equation, were used to determine the differential enthalpy, differential entropy, and Gibbs free energy present in the process. In addition, the compensation theory for this system was validated. The results indicate an overall reduction in the energy required to establish equilibrium in the water-seed system as the moisture content increases. A relationship is observed that will inject Gibbs free energy into the system at the extremes of the magnetic dose applied to the seeds. This relationship is stable as a first approximation to understand the energetic aspects when connecting the magnetic processing of seeds.

## 1. Introduction

Magnetic seed treatment (MST) is a technique that has been in development since its first approaches in the early 60s with two studies that showed for the first time an application of magnetic field in the improvement of tomato[1] and wheat[2] production. Today it is known that these techniques that involve the use of magnetism to stimulate seeds have potential use in the improvement of germination-based processes, establishment of seedlings and in general the improvement of crop production. The application and favorable use of MST has been compiled along the last years in several reviews [3]–[5] always pointing the two main ways of studying the effect of magnetic fields in seeds and foodstuff: the improvement of plant structures after the exposure and the identification of changes in the biochemical and biophysical processes of the plant.

Many studies regarding the biophysical characterization of processes in seeds are based in the experimental analysis of sorption behavior in different environmental conditions, some of the studies that can be identified in the literature include as samples different seeds like pinhao [6]–[9], pickly pear[10], watermelon[11], pumpkin[12], beans[13]. These studies usually give a thermodynamic panorama of the adsorption process, involving parameters like temperature or chemical treatments to the samples but no the enrolling of magnetic field.

Despite the fact the magnetic seed treatment has been widely studied for years now, there is a lack of understanding on the modified processes that allow magnetically seeds treatment to develop a better rate of germination and growth[14].

The aim of this study is to give a qualitative notion of the relation between the energy of the overall system implied in the adsorption process and the magnetic treatment applied to the samples, in order to do this different models from the literature were fitted to the data and the most convenient was selected in order to apply the analysis. Different parameters including free energy were obtained from the model and a tendency between free energy and magnetic dose was determined.

## 2. Materials and methods

Samples, magnetic treatments and experimentally obtained adsorption isotherms implied in this study were retrieved from a previous study of the same author [15] and main data is pointed in Table 1.

**Table 1.**

|  |  | Equilibrium moisture content | | | | |
|---|---|---|---|---|---|---|
| Salt | Water activity ($a_W$) | Control | $T_1$[mT] | $T_2$[mT] | $T_3$[mT] | $T_4$[mT] |
| $CH_3CO_2K$ | 2.16E-01 | 0.0547 | 0.0562 | 0.0511 | 0.0544 | 0.0553 |
| $MgCl_2$ | 3.24E-01 | 0.0728 | 0.0797 | 0.0692 | 0.087 | 0.0814 |
| $K_2CO_3$ | 4.32E-01 | 0.0728 | 0.074 | 0.0664 | 0.0654 | 0.0849 |
| $NaBr$ | 5.60E-01 | 0.0755 | 0.0812 | 0.0785 | 0.076 | 0.0773 |
| $KI$ | 6.79E-01 | 0.1082 | 0.0997 | 0.1151 | 0.1031 | 0.1078 |
| $NaCl$ | 7.51E-01 | 0.1209 | 0.1182 | 0.1094 | 0.1073 | 0.1052 |
| $KCl$ | 8.36E-01 | 0.1888 | 0.1057 | 0.1446 | 0.106 | 0.1061 |
| $K_2SO_4$ | 9.70E-01 | 0.3079 | 0.2885 | 0.2733 | 0.261 | 0.2404 |

### 2.1. Samples

For the samples, maize seeds (*Zea mays L.* cv. ICAV305, Semillas del Pacífico, Cartago, Colombia) were sieved and selected to obtain samples with $0.3878\ g\ \pm\ 0.0002\ g$ average mass and $0.356\ cm^3\ \pm\ 0.008\ cm^3$ average volume. Initial moisture content ($M_0$) was obtained by gravimetric technique yielding to an initial average moisture content of 15.1% [16].

### 2.2. Magnetic treatment

The magnetic field was generated using a circular coil electromagnet (5403, GMW, San Carlo, CA), fed with direct current (DC) source (N5768A, Agilent Technologies, Santa Clara, CA).

The control is the experiment carried over samples with no magnetic treatments. Besides control, four treatments were carried over the samples: $T_1, T_2, T_3$ and $T_4$ with an exposure time of $10\ min$ and magnetic induction values of $80, 120, 160$ and $200\ mT$ respectively. The homogeneity of magnetic induction was 98.4%. The values used in the magnetic treatment can be better described by using the concept of magnetic flux density for a static magnetic field[17]:

$$D = \frac{t}{2\mu_0} B^2$$

Where $t$ is the time of exposure (10 minutes in every case), $\mu_0$ is the vaccum magnetic permeability $\mu_0 = 4\pi \times 10^{-7} H/m$ and $B$ is the magnetic induction. The values of magnetic dose for $80, 120, 160$ and $200\ mT$ are $1.528, 3.438, 6.112$ and $9.549\ kJ \times s/cm^3$ respectively.

### 2.3. Adsorption Isotherms

Samples used in the adsorption process were containers with a cylindrical shape and a porous surface that allowed the storage of 75 seeds each. In order to stablish different pressure values in the system, saturated salt solutions were used allowing a range of water activities from $0.216\ to\ 0.970$[18] according to the table 1.

Saturated solutions were prepared taking into account solubility of the salts at a temperature of $30\ °C$. The samples were placed into pertri dishes next to saturated solutions prepared with 20 ml of each solution. The system was set to 30 °C and every experiment was carried out in triplicate following the gravimetric standard method [19] with the use of an analytical balance in order to obtain the equilibrium moisture content (EMC), further details regarding the experimental setup can be found in the paper[15].

### 2.4. Data Analysis

The experimental adsorption data obtained from a previous work was fitted to ten moisture sorption isotherm models (table 2) for every magnetic treatment. The chosen models are a mix between temperature dependent and temperature independent models, both were used in order to obtain the best fit of the overall experiment. The fit were computed using the free software CurveExpertBasic. Brunauer-Emmet-Teller (BET) [20] model was not tested because it is in acceptable agreement with isotherm curves only for water activities $a_W < 0.5$ [21], while the range of water activities used for the experiments is wider. In order to analyze the reliability of the models the coefficient of determination ($r^2$) and the standard error were identified for each case. As for $r^2$ a value closer to 1 will give a better agreement between the experimental points and the non-linear fit, a lower value of standard error will provide a better agreement.

### 2.4.1. Thermodynamic parameters

In order to determine the relation between the free energy and the magnetic induction it is necessary to find the values of differential enthalpy, it can be determined from both experimental and fitted data using an equation derived from the Clausius-Clayperon equation[22]

$$\Delta h = -R \left[ \frac{\partial \ln(a_W)}{\partial \left(\frac{1}{T}\right)} \right]_X$$

where $a_W$ is the water activity, $T$ is the absolute temperature $[K]$, $R$ is the ideal gas constant $R = 8.314472 \frac{J}{Kmol}$, $X$ indicates a constant moisture content $\left[\frac{kg\ water}{kg\ dry\ solid}\right]$ and $\Delta h$ is the differential enthalpy $\left[\frac{J}{mol}\right]$.

The differential enthalpy of sorption was calculated plotting $\ln(a_W)$ against $\frac{1}{T}$ for constant moisture content. The slope of the linear regression is equal to $-\frac{\Delta h}{R}$. This procedure was repeated for different moisture contents in order to evidence the dependence of differential enthalpy with moisture content. In the aim of obtaining thermodynamic data from the experiment, several temperatures must be used the calculation process, as this particular experiment was carried in only one temperature (30°C), the temperature-dependent model (that allowed us to predict the isothermal behavior at different temperatures) with higher $r^2$ in control was selected. After the selection of the model, it was used in order to obtain water activities $a_W$ for each magnetic treatment at different moisture contents under the assumption that differential enthalpy is invariant with temperature [22].

The differential enthalpy $\Delta h$ and differential entropy $\Delta S$ are related by the equation

$$\ln(a_W)_X = -\frac{\Delta h}{RT} + \frac{\Delta S}{R}$$

so the differential entropy can be obtained from the linear coefficient $-\frac{\Delta S}{R}$ using the linear regression of $\ln(a_W)$ against $\frac{1}{T}$ for a constant moisture content [23].

It is possible to relate all the thermodynamic parameters through compensation theory [24], [25]. This theory gives a linear relationship between $\Delta h$ and $\Delta S$ through the equation

$$\Delta h = T_B \Delta S + \Delta G_B$$

where $T_B$ $[K]$ es the isokinetic temperature and $\Delta G_B$ is the free energy $[J/mol]$ calculated by linear regression. From thermodynamic theory, free energy gives an insight in the affinity of the sorbent for water and provides information about the spontaneousness of the process, being spontaneous ($\Delta G < 0$) or non-spontaneous ($\Delta G > 0$) [26].

With the values of $\Delta G_B$ determined, it is possible to do a statistical corroboration of the compensation theory [27] by the comparison of $T_B$ and the harmonic mean temperature $T_{hm}$ [$K$]

$$T_{hm} = \frac{n}{\sum_1^n \left(\frac{1}{T}\right)}$$

with n being the number of isotherms used.

Compensation theory only applies if $T_B \neq T_{hm}$.

## 3. Results and discussion

Figure 1 shows the experimental values of equilibrium moisture content of maize as a function of $a_W$ for several magnetic treatments. $a_W$ was in the range of $0.216 < a_W < 0.969$ due to the saturated salts used in the experiment. In every case it is possible to observe the tendency of the equilibrium moisture content to be higher with higher water activities for every treatment. This growth relation is general for every absorbent studied in the literature even if it is not food related. In this particular case the curve follows the type – II shape in BET classification [20] that has been observed in several foodstuff such as potato starch, bananas, onion, carrot, green pepper, sesame seeds and many more [28].

It can be seen in figure 1 that in general it exists a decrease of the equilibrium moisture content with the increase of the magnetic dose applied to the samples, mostly at higher water activities $(a_W > 0.75)$ as pointed before by Vashitsh et al [29]. Even though the EMC for every treatment is lower than the one for control in the whole range of $a_W$ the increase or decrease between consecutive increasing treatments seems to be spread randomly. It is possible to assume that there is no a linear relation between magnetic dose and EMC but the determination of this relations is out of the scope of this work.

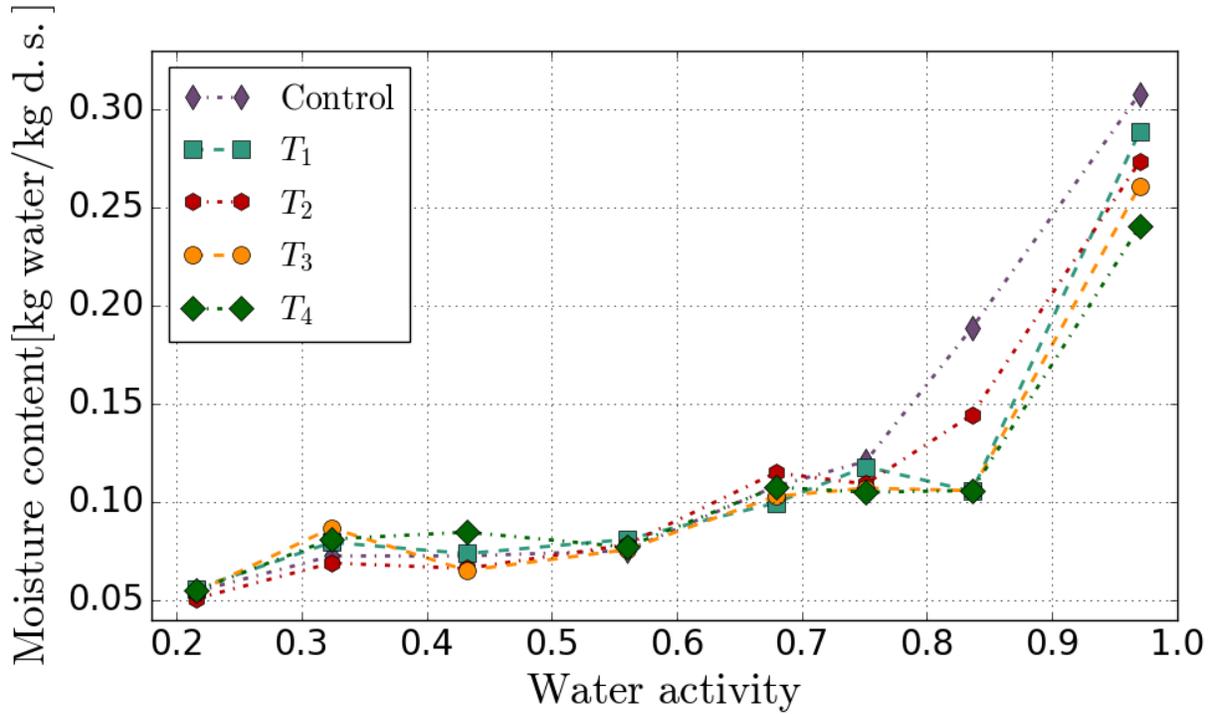

**Figure 1**

The values of the isotherms were fitted to the models listed in Table 2. Table 3 shows the values of $r^2$ and standard error for each model. In control and every treatment the Peleg[30] model was the one with the best value of $r^2$ being 0.9964 for control, 0.9943 for $T_1$, 0.9944 for $T_2$, 0.9873 for $T_3$ and 0.9873 for $T_4$. In every case sorption models were poorly fitted for $T_3$ and $T_4$ having $r^2$ below 0.99. Despite the poor fitting for the last treatments, the information regarding modeling indicates the Peleg as the best model for this particular study considering the temperature at which the experiment was carried on.

**Table 2**

| | |
|---|---|
| Oswin [31] | $M = c\left(\dfrac{x}{1-x}\right)^n$ |
| Peleg [30] | $M = x^b + dx^e$ |
| GAB[32] | $M = \dfrac{mckx}{(1-(kx))(1-kx+ckx)}$ |
| Smith[33] | $M = c + b\ln(1-x)$ |
| Halsey[34] | $M = \left(-\dfrac{a}{\ln(x)}\right)^{\left(\frac{1}{n}\right)}$ |
| Henderson[35] | $M = \left(-\left(\dfrac{\ln(1-x)}{c}\right)\right)^{\frac{1}{n}}$ |
| Modified Chung-Pfost [36], [37] | $M = \left(-\dfrac{1}{c}\right)\ln\left(\dfrac{-(T+c)*\ln(x)}{a}\right)$ |
| Modified Halsey [36], [37] | $M = \left(\dfrac{-\exp(a+bT)}{\ln(x)}\right)^{\frac{1}{c}}$ |
| Modified Oswin [36], [37] | $M = (a+bT)\left(\dfrac{1-x}{x}\right)^{-\frac{1}{c}}$ |
| Modified Henderson [36], [37] | $M = \left(\dfrac{\ln(1-x)}{-a(T+b)}\right)\left(\dfrac{1}{c}\right)$ |

**Table 3**

| | Control | | $T_1$ | | $T_2$ | | $T_3$ | | $T_4$ | |
|---|---|---|---|---|---|---|---|---|---|---|
| Model | $r^2$ | Stand err | $r^2$ | Stand err | $r^2$ | Stand err | $r^2$ | Stand err | $r^2$ | Stand err |
| Oswin | 0.9847 | 0.0160 | 0.9751 | 0.0176 | 0.9948 | 0.0078 | 0.9702 | 0.0170 | 0.9702 | 0.0170 |
| Peleg | 0.9964 | 0.0096 | 0.9943 | 0.0104 | 0.9944 | 0.0100 | 0.9873 | 0.0136 | 0.9873 | 0.0136 |
| GAB | 0.9925 | 0.0123 | 0.9715 | 0.0206 | 0.9935 | 0.0096 | 0.9659 | 0.0199 | 0.9659 | 0.0199 |
| Smith | 0.9875 | 0.0145 | 0.9544 | 0.0237 | 0.9914 | 0.0101 | 0.9541 | 0.0210 | 0.9541 | 0.0210 |
| Halsey | 0.9812 | 0.0178 | 0.9834 | 0.0144 | 0.9934 | 0.0088 | 0.9789 | 0.0143 | 0.9789 | 0.0143 |
| Henderson | 0.9799 | 0.0183 | 0.9300 | 0.0292 | 0.9792 | 0.0157 | 0.9233 | 0.0270 | 0.9233 | 0.0270 |
| Modified Chung-Pfost | 0.9747 | 0.0205 | 0.9295 | 0.0293 | 0.9786 | 0.0159 | 0.9301 | 0.0258 | 0.9301 | 0.0258 |
| Modified Halsey | 0.9812 | 0.0194 | 0.9834 | 0.0158 | 0.9934 | 0.0097 | 0.8824 | 0.0362 | 0.8824 | 0.0362 |
| Modified Oswin | 0.9847 | 0.0175 | 0.9751 | 0.0193 | 0.9948 | 0.0086 | 0.9702 | 0.0186 | 0.9702 | 0.0186 |
| Modified Henderson | 0.9799 | 0.0201 | 0.9300 | 0.0320 | 0.9792 | 0.0172 | 0.9233 | 0.0296 | 0.9233 | 0.0296 |

### 3.1. Thermodynamic parameters

In order to calculate the differential enthalpy, experimental data was fitted to a temperature dependent model for the sake of temperature predictions which are necessary for the analysis [22]. The data was fitted to Modified Henderson model for having the higher $r^2$ value between the temperature predictive models for control and every treatment independently. The capacity of modulate the temperature is necessary in order to determine thermodynamic parameters. Modified Henderson model is described for every magnetic treatment as

$$M = \left[\left(\frac{ln(1-a_W)}{-a(T+b)}\right)\frac{1}{c}\right]_B$$

The equation was used to determine water activities for constant moisture contents at different temperatures. Fitted equation for experimental temperature T=303 K and different predicted values (T=298 K and T=308 K) for control are presented in figure 2. Several water activity values where then used in Clausius-Clapeyron equation to obtain differential enthalpy values at different moisture contents for all the treatments.

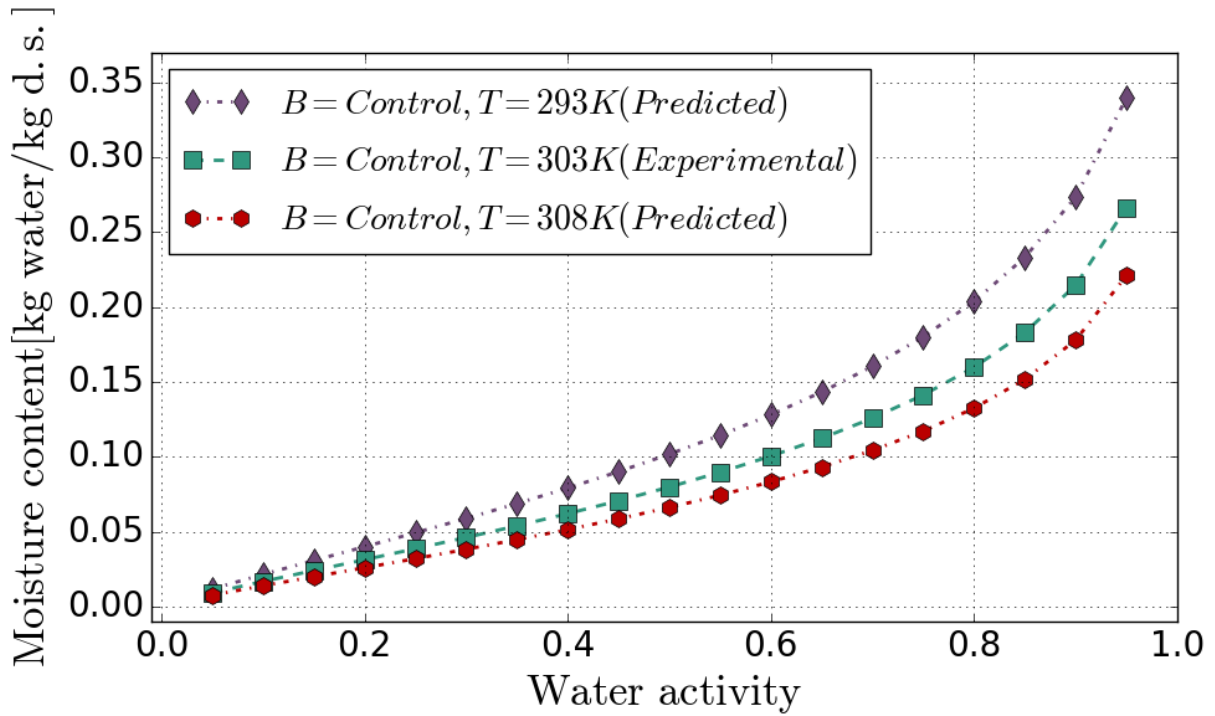

**Figure 2**

Figure 3 shows that the differential enthalpy is highly dependent on moisture content. There is a higher binding energy for removal of water at lower values of moisture sorption $M < 0.13$ but it decreases as moisture content increases. The differential enthalpy converging to zero at higher moisture contents suggests moisture existing in free form for high moisture content values[7] due to the lack of strong binding sites between the water and the surface after the fulfilling of the monolayer. A similar behavior has been observed in previous studies in food products such as tamarind[38], lettuce[39], barley[40] and potato[23].

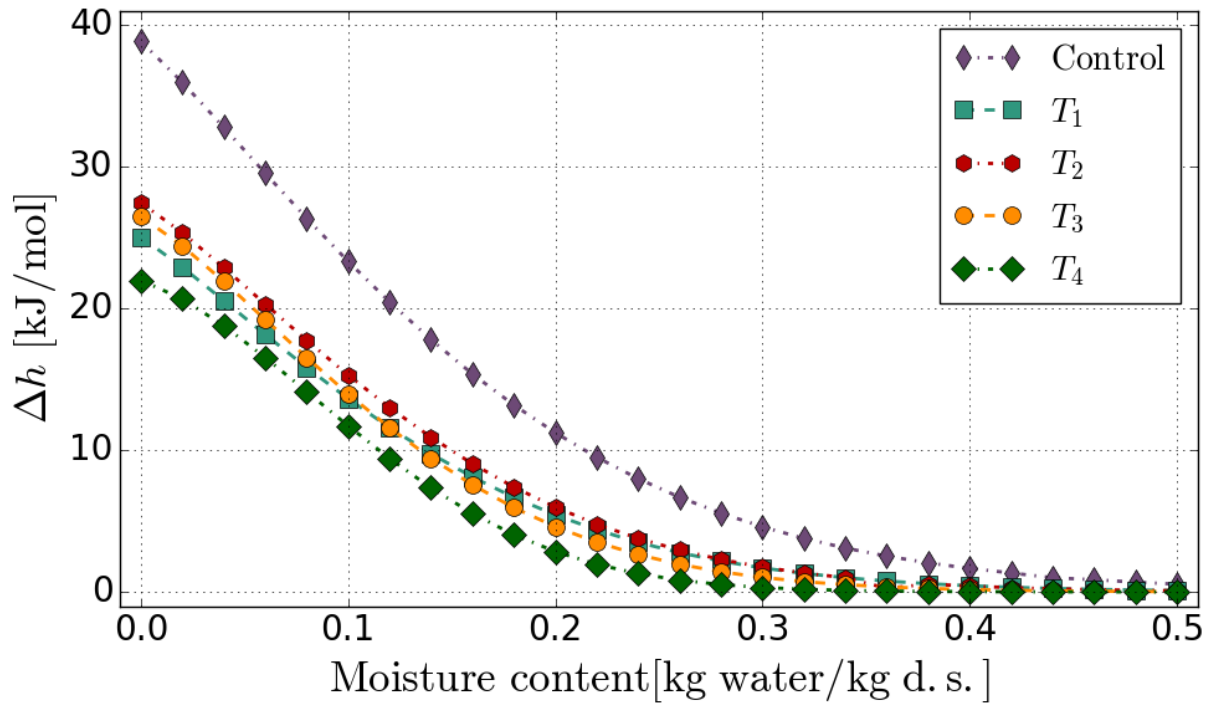

**Figure 3**

Even though the behavior of the differential enthalpy with the change of moisture content is the same for every magnetic treatment, it is clear that magnetically treated seeds show lower values of differential enthalpy for every moisture content which indicates the decrease of the binding energy of water in the surface of the sample with higher magnetic doses. Exploring the evolution of enthalpy with different magnetic treatments, Figure 3 shows that as a lower magnetic induction ($T1$) results in a lower differential enthalpy than a higher magnetic induction ($T2, T3$) even though an even higher dose ($T4$) shows the lowest enthalpy for every for every EMC value. This behavior exposes the non-linear dependence of the enthalpy with the increase of magnetic dose. This relation between magnetic treatment and differential enthalpy allows to bring back previous results regarding the non-linearity between biological mechanisms of plants such as germination and magnetic dose [15], [29].

Figure 4 shows the behavior of differential entropy ($\Delta S$) with the variation of moisture content $M$ for different treatments. The tendency keeps similar to the one shown by the differential enthalpy, for every treatment the differential entropy has higher values at lower equilibrium moisture content ($M < 0.13$) and decreases until it converges to zero. An increase in $\Delta S$ can be seen for the second data point analyzed in every treatment ($M = 0.03$), from there on the behavior keeps the same as $\Delta h$ showing a fast decrease until it reaches zero.

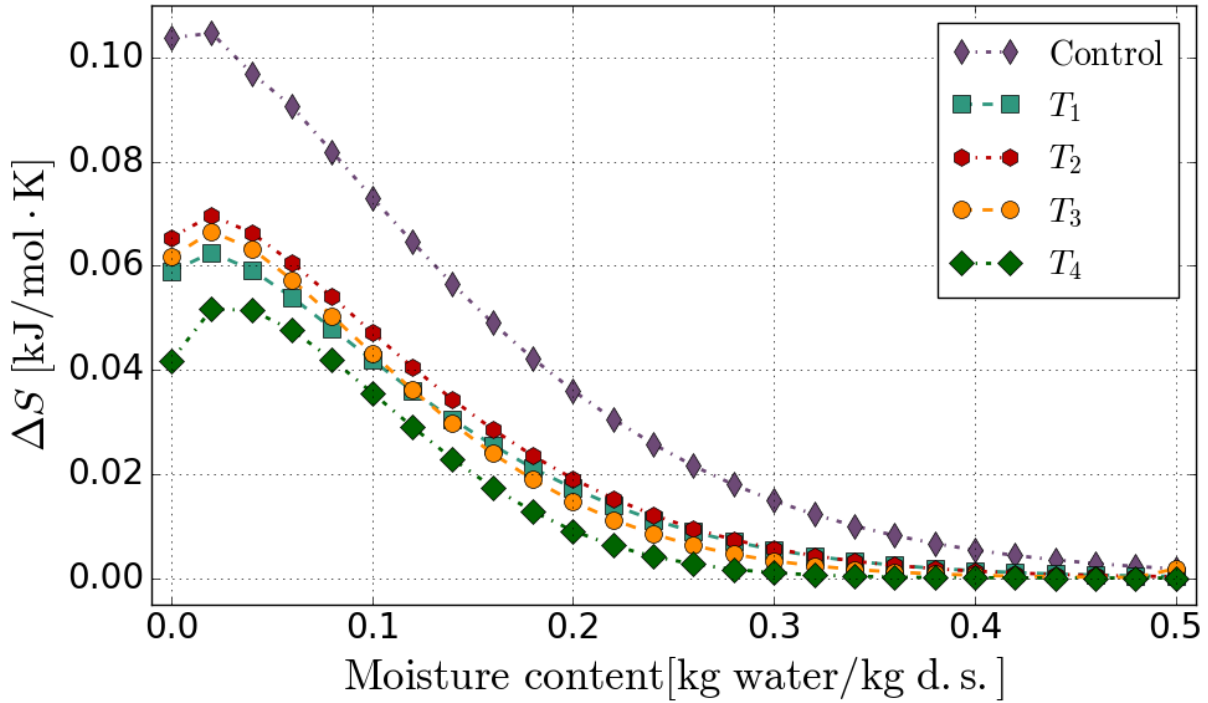

**Figure 4**

The initial increase in $\Delta S$ (figure 4) can be attributed to the fact that entropy gives a reason of the mobility of the water molecules, with low moisture content there is an interaction between two mechanisms[41]: the first mechanism involves the system composed by the free water and the surface, as there is no bonded water, there are not multimolecular interactions (interactions between free and bonded water) so the molecules are not likely to move a lot. The second mechanism that involves the lack of adsorbed molecules drives to a lot a free space for the molecules to move along the system. The maximum value of entropy found around $M = 0.15$ could be possible due to the equilibrium between the mechanisms where there are some water molecules adsorbed by the surface and also enough space for the free water to move.

Figure 5 shows the differential enthalpy ($\Delta h$) as a function of differential entropy ($\Delta S$) for every treatment. This relation is useful in order to validate the compensation theory which allows the determination of the system's free energy. The linear relation between these variables was obtained (as indicated by the determination coefficients, Table 4) validating the compensation of the system for each treatment. The parameters of isokinetic temperature ($T_B$) and free energy ($\Delta G$) were obtained by linear regression. $T_B$ is the temperature at which all the sorption reactions take place at the same rate. Five different values were obtained, one for each treatment as pointed in Table 4. The values slightly differ from each other being higher at higher magnetic induction($382.04\ K$ for $B = 200\ mT$) and lower for control ($342.58\ K$). Isokinetic temperature ($T_{hm}$) was obtained with a value of $264.36\ K$ differing from all $T_B$ which validates, once again, the isokinetic theory[24].

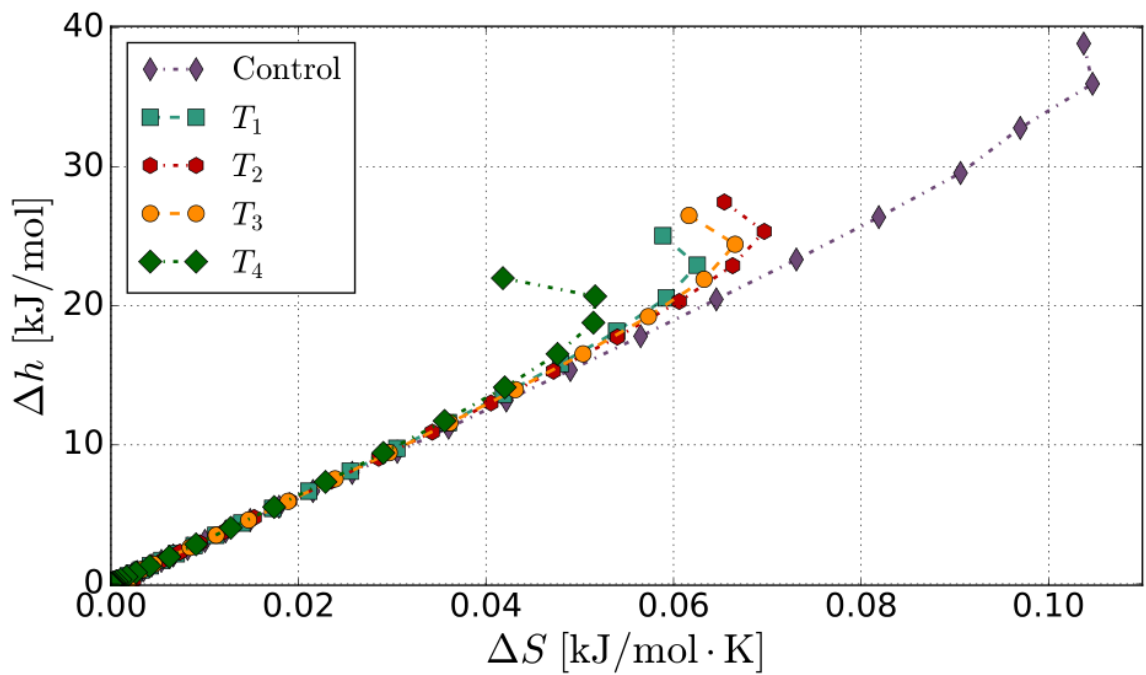

**Figure 5**

**Table 4**

|  | $T_B [K]$ | $\Delta G \left[\frac{kJ}{mol}\right]$ | $r^2$ |
|---|---|---|---|
| Control | 342.58 | 0.5542 | 0.9934 |
| T1 | 361.72 | 0.3798 | 0.9838 |
| T2 | 358.97 | 0.4207 | 0.9851 |
| T3 | 362.03 | 0.3687 | 0.9832 |
| T4 | 382.04 | 0.2374 | 0.96 |

According to Leffer theory[24], for the case of $T_{hm} < T_B$, the process is enthalpy-driven, which is the case in this study for every treatment, opposite to a entropy-driven process ($T_{hm} < T_B$). Information on the enthalpy–entropy compensation could be an important tool to recognize different mechanisms for water sorption under varying conditions[42].

The values obtained from linear regression for free energy ($\Delta G$) are always positive $\Delta G > 0$ which suggest a non-spontaneous reaction process. The free energy is an indicative of the amount of energy needed for making sorption sites available [43].

Figure 6 shows a relation between the amount of free energy ($\Delta G$) and the amount of magnetic dose for control and every treatment, even though the magnetic dose is dependent on several variables, a first look on the direct dependence can be seen for $\Delta G$. For the dependence of $\Delta G$ with $D$ a straight line was drawn over the data in order to identify the tendency. This tendency allows us to determine that a higher magnetic dose will impact in the free energy lowering its values. In the specific case regarding the adsorption process, it could possibly indicate that a higher magnetic dose applied to the seed samples allows them to reach moisture equilibrium with a lower amount of energy applied to the system. This result is in agreement with the conclusions obtained by [15] where the D'arcy-Watt model [44] showed an increase of adsorption sites with the exposure to higher magnetic dose.

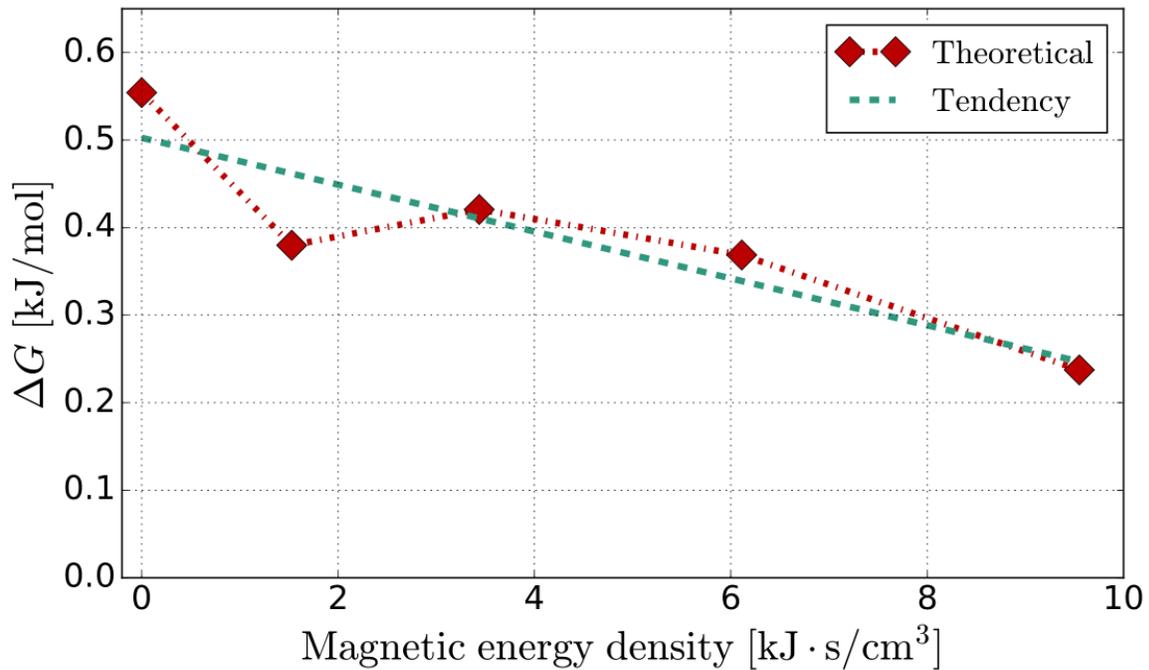

**Figure 6**

Even though different models give reason of different parameters regarding the adsorption processes, the agreement between several theoretical and experimental results converge into a general improvement in the process despite the studied adsorption feature once the seeds are exposed to static magnetic fields.

## 4. Conclusions

Adsorption isotherms of magnetically treated maize seeds obtained from a previous work were analyzed in order to determine their thermodynamic properties. The data obtained at 303 K was successfully fitted to the Modified-Henderson model in order to predict the adsorption isotherms at 298 K and 308 K for the determination of thermodynamic parameters.

Differential enthalpy ($\Delta h$) and differential entropy ($\Delta S$) changes were determined with the increasement of moisture content. Values of $\Delta h$ decrease from a maximum value of $38.8618 \; kJ/mol$ for control and in general lower values for every treatment. The values of $\Delta S$ also decreases after a local increase of $0.1047 \; kJ/mol \cdot K$ with the increase of moisture content.

Compensation theory was validated indicating an enthalpy driven reaction and also the free energy ($\Delta G$) was determined for every magnetic treatment showing a decrease with the increase of magnetic dose applied to the system, validating thermodynamically the improvement of the adsorption process with the exposure of the sample to different magnetic doses.

## 5. References

[1]   A. A. BOE and D. K. SALUNKHE, "Effects of Magnetic Fields on Tomato Ripening," *Nature*, vol. 199, no. 4888, pp. 91–92, Jul. 1963.

[2]   U. J. Pittman, "GROWTH REACTION AND MAGNETOTROPISM IN ROOTS OF WINTER WHEAT (KHARKOV 22 M.C.)," *Can. J. Plant Sci.*, vol. 42, no. 3, pp. 430–436, Jul. 1962.

[3]   J. A. Teixeira da Silva and J. Dobránszki, "Magnetic fields: how is plant growth and development impacted?," *Protoplasma*, vol. 253, no. 2, pp. 231–248, 2016.

[4]   S. Pietruszewski and E. Martínez, "Magnetic field as a method of improving the quality of sowing material: a review," *Int. Agrophysics*, vol. 29, no. 3, pp. 377–389, Jul. 2015.

[5]   M. E. Maffei, "Magnetic field effects on plant growth, development, and evolution," *Front. Plant Sci.*, vol. 5, no. September, pp. 1–15, 2014.

[6]   F. Cladera-Olivera *et al.*, "Water sorption isotherms of fresh and partially osmotic dehydrated pumpkin parenchyma and seeds at several temperatures," *Int. J. Food Sci. Technol.*, vol. 24, no. 1, pp. 1257–1264, 2010.

[7]   F. Cladera-Olivera, A. C. Pettermann, C. P. Z. Noreña, K. Wada, and L. D. F. Marczak, "Thermodynamic properties of moisture desorption of raw pinhão (Araucaria angustifolia seeds)," *Int. J. Food Sci. Technol.*, vol. 43, no. 5, pp. 900–907, 2008.

[8]   F. CLADERA-OLIVERA, L. D. F. MARCZAK, C. P. Z. NOREÑA, and A. C. PETTERMANN, "Modelling water adsorption isotherms of pinhao flour and thermodynamic analysis of the adsorption process," *J. Food Process Eng.*, vol. 34, no. 3, pp. 826–843, Jun. 2011.

[9] J. C. Spada, C. P. Z. Noreña, L. D. F. Marczak, and I. C. Tessaro, "Water adsorption isotherms of microcapsules with hydrolyzed pinhão (Araucaria angustifolia seeds) starch as wall material," *J. Food Eng.*, vol. 114, no. 1, pp. 64–69, 2013.

[10] L. Hassini, E. Bettaieb, H. Desmorieux, S. S. Torres, and A. Touil, "Desorption isotherms and thermodynamic properties of prickly pear seeds," *Ind. Crops Prod.*, vol. 67, pp. 457–465, 2015.

[11] A. A. Wani, D. S. Sogi, U. S. Shivhare, I. Ahmed, and D. Kaur, "Moisture adsorption isotherms of watermelon seed and kernels," *Dry. Technol.*, vol. 24, no. 1, pp. 99–104, 2006.

[12] L. Mayor, R. Moreira, F. Chenlo, and A. M. Sereno, "Water sorption isotherms of fresh and partially osmotic dehydrated pumpkin parenchyma and seeds at several temperatures," *Eur. Food Res. Technol.*, vol. 220, no. 2, pp. 163–167, 2005.

[13] Y. Oladosu *et al.*, "Principle and application of plant mutagenesis in crop improvement: A review," *Biotechnol. Biotechnol. Equip.*, vol. 30, no. 1, pp. 1–16, 2016.

[14] P. Galland and A. Pazur, "Magnetoreception in plants," *J. Plant Res.*, vol. 118, no. 6, pp. 371–389, 2005.

[15] J. I. Torres-Osorio, J. E. Aranzazu-Osorio, and M. V. Carbonell-Padrino, "Efecto del campo magnético estático homogéneo en la germinación y absorción de agua en semillas de soja," *Tecno Lógicas*, vol. 18, no. 35, pp. 11–20, 2015.

[16] A. International, *AOAC: Official Methods of Analysis, 2016*, vol. 552, no. c. 2016.

[17] S. Pietruszewski and E. Martínez, "Magnetic field as a method of improving the quality of sowing material: A review," *Int. Agrophysics*, vol. 29, no. 3, pp. 377–389, 2015.

[18] L. Greenspan, "Humidity fixed points of binary saturated aqueous solutions," *J. Res. Natl. Bur. Stand. Sect. A Phys. Chem.*, vol. 81A, no. 1, p. 89, Jan. 1977.

[19] A. International, *Official Methods of Analysis of the Association of Official Analytical Chemists*, vol. 60, no. 2. 1971.

[20] S. Brunauer, P. H. Emmett, and E. Teller, "Gases in Multimolecular Layers," *J. Am. Chem. Soc.*, vol. 60, no. 1, pp. 309–319, 1938.

[21] K. Jin Park, Z. Vohnikova, and F. Pedro Reis Brod, "Evaluation of drying parameters and desorption isotherms of garden mint leaves (Mentha crispa L.)," *J. Food Eng.*, vol. 51, no. 3, pp. 193–199, 2002.

[22] E. Tsami, "Net isoteric heat of sorption in dried fruits," *J. Food Eng.*, vol. 14, no. 4, pp. 327–335, 1991.

[23] W. A. M. McMinn and T. R. A. Magee, "Thermodynamic properties of moisture sorption of potato," *J. Food Eng.*, vol. 60, no. 2, pp. 157–165, 2003.

[24] E. Grunwald and J. E. Leffler, "Rates and Equilibria of Organic Reactions: As Treated by Statistical, Thermodynamic, and Extrathermodynamic Methods.," 1963.

[25] V. R. . Telis, A. . Gabas, F. . Menegalli, and J. Telis-Romero, "Water sorption thermodynamic

properties applied to persimmon skin and pulp," *Thermochim. Acta*, vol. 343, no. 1–2, pp. 49–56, 2002.

[26] D. APOSTOLOPOULOS and S. G. GILBERT, "Water Sorption of Coffee Solubles by Frontal Inverse Gas Chromatography: Thermodynamic Considerations," *J. Food Sci.*, vol. 55, no. 2, pp. 475–487, 1990.

[27] R. R. Krug, W. G. Hunter, and R. A. Grieger, "Enthalpy-entropy compensation. 2. Separation of the chemical from the statistical effect," *J. Phys. Chem.*, vol. 80, no. 21, pp. 2341–2351, 1976.

[28] S. Yanniotis and J. Blahovec, "Model analysis of sorption isotherms," *LWT - Food Sci. Technol.*, vol. 42, no. 10, pp. 1688–1695, 2009.

[29] A. Vashisth and S. Nagarajan, "Characterization of water binding and germination traits of magnetically exposed maize (Zea mays L.) seeds equilibrated at different relative humidities at two temperatures," *Indian J. Biochem. Biophys.*, vol. 46, no. 2, pp. 184–191, 2009.

[30] M. PELEG, "ASSESSMENT of A SEMI-EMPIRICAL FOUR PARAMETER GENERAL MODEL FOR SIGMOID MOISTURE SORPTION ISOTHERMS," *J. Food Process Eng.*, vol. 16, no. 1, pp. 21–37, 1993.

[31] C. R. Oswin, "The kinetics of package life. III. The isotherm," *J. Soc. Chem. Ind.*, vol. 65, no. 12, pp. 419–421, Dec. 1946.

[32] E. A. Guggenheim, *Application of statistical mechanics*. Oxford Press, 1966.

[33] S. E. Smith and S. E. Smith, "The Sorption of Water Vapor by High Polymers," *J. Am. Chem. Soc.*, vol. 69, no. 3, pp. 646–651, 1947.

[34] G. Halsey, "Physical Adsorption on Non-Uniform Surfaces," *J. Chem. Phys.*, vol. 16, no. 10, pp. 931–937, Oct. 1948.

[35] S. Henderson, "A basic concept of equilibrium moisture," *Agric. Eng.*, no. 33, pp. 29–32.

[36] T. L. Thompson and R. M. Peart and G. H. Foster, "Mathematical Simulation of Corn Drying —A New Model," *Trans. ASAE*, vol. 11, no. 4, pp. 0582–0586, 2013.

[37] A. Standars, "ASAE D245.6.15 DEC02: Moisturerelationships of plant-based agricultural products.," 2007.

[38] E. Alpizar-Reyes, H. Carrillo-Navas, R. Romero-Romero, V. Varela-Guerrero, J. Alvarez-Ramírez, and C. Pérez-Alonso, "Thermodynamic sorption properties and glass transition temperature of tamarind seed mucilage (Tamarindus indica L.)," *Food Bioprod. Process.*, vol. 101, pp. 166–176, 2017.

[39] J. S. Zeymer, P. C. Corrêa, G. H. H. De Oliveira, and F. M. Baptestini, "Thermodynamic properties of water desorption in lettuce seeds," *Semin. Agrar.*, vol. 39, no. 3, pp. 921–931, 2018.

[40] M. R. Cruz, A. M. D. Canteli, F. A. P. Voll, L. C. B. Zuge, and A. de P. Scheer, "Statistical evaluation of models for sorption and desorption isotherms for barleys," *Acta Sci. - Technol.*, vol. 40, no. 1986, 2018.

[41] J. D. Hoyos-Leyva, L. A. Bello-Pérez, and J. Alvarez-Ramirez, "Thermodynamic criteria analysis for the use of taro starch spherical aggregates as microencapsulant matrix," *Food Chem.*, vol. 259, no. March, pp. 175–180, 2018.

[42] C. I. Beristain, H. S. Garcia, and E. Azuara, "Enthalpy-Entropy compensation in food vapor adsorption," *J. Food Eng.*, vol. 30, no. 3–4, pp. 405–415, Nov. 1996.

[43] Y. N. Nkolo Meze'e, J. Noah Ngamveng, and S. Bardet, "Effect of enthalpy-entropy compensation during sorption of water vapour in tropical woods: The case of Bubinga (Guibourtia Tessmanii J. Léonard; G. Pellegriniana J.L.)," *Thermochim. Acta*, vol. 468, no. 1–2, pp. 1–5, 2008.

[44] R. L. D'Arcy and I. C. Watt, "Analysis of sorption isotherms of non-homogeneous sorbents," *Trans. Faraday Soc.*, vol. 66, p. 1236, 1970.